\documentclass[aps,prb,twocolumn,groupedaddress]{revtex4}

\usepackage{amsmath}
\usepackage{graphicx}

\def\Vec#1{\mbox{\boldmath $#1$}}

\begin{document}

\title{Improved chain mean-field theory for quasi-one-dimensional
  quantum magnets}

\author{Synge Todo$^{1,2}$}
\email{wistaria@ap.t.u-tokyo.ac.jp}
\author{Akira Shibasaki$^1$}
\thanks{Present address: Canon Inc., Tokyo 146-8501, Japan}
\affiliation{$^1$Department of Applied Physics, University of Tokyo, Tokyo
  113-8656, Japan}
\affiliation{$^2$CREST, Japan Science and Technology Agency, Kawaguchi
  332-0012, Japan}
\date{\today}

\begin{abstract}
  A novel mean-field approximation for quasi-one-dimensional (Q1D)
  quantum magnets is formulated.  Our new mean-field approach is based
  on the Bethe-type effective-field theory, where thermal and quantum
  fluctuations between the nearest-neighbor chains as well as those in
  each chain are taken into account exactly.  The self-consistent
  equation for the critical temperature contains the boundary-field
  magnetic susceptibilities of a multichain cluster, which can be
  evaluated accurately by some analytic or numerical methods, such as
  the powerful quantum Monte Carlo method.  We show that the accuracy
  of the critical temperature of Q1D magnets as a function of the
  strength of interchain coupling is significantly improved, compared
  with the conventional chain mean-field theory.  It is also
  demonstrated that our new approximation can predict nontrivial
  dependence of critical temperature on the sign (i.e., ferromagnetic
  or antiferromagnetic) of interchain coupling as well as on the
  impurity concentration in randomly diluted Q1D Heisenberg
  antiferromagnets.
\end{abstract}


\maketitle

\section{Introduction}

Space dimensionality plays an essential role in phase transitions and
critical phenomena of quantum magnets.  As the dimension is lowered,
effects of thermal and quantum fluctuations generally become stronger.
As a result, the quantum antiferromagnetic Heisenberg model in two
dimensions, for example, does not exhibit long-range order any more
except at the ground state,\cite{MerminW1966} though a
finite-temperature phase transition occurs in its three-dimensional
counterpart.  Furthermore, there is no long-range order even at zero
temperature in one dimension.\cite{Bethe1931}  Such one-dimensional
magnets arose great interests as many novel phenomena characteristic
to systems with strong quantum fluctuations, e.g., Tomonaga-Luttinger
liquid state or Haldane gap state, have been observed theoretically as
well as in the experiments.\cite{SchollwoeckRFB2004}

Real materials, however, can not be purely one dimensional but three
dimensional, i.e., there always exist interactions between
one-dimensional chains ({\em interchain} interactions) albeit it is
much weaker, often by orders of magnitude, than the dominant
interactions along the chains ({\em intrachain} interactions).
Three-dimensional materials with strong spatial anisotropy are often
referred to as quasi-one-dimensional (Q1D) systems.  Indeed, in many
Q1D materials a long-range order emerges at low temperatures, which is
a direct consequence of three dimensionality of the system.  In order
to explain low-energy behavior of such Q1D materials correctly, a
theory which properly incorporates the effect of interchain
interaction is essential.

So far, effects of weak interchain interactions in Q1D quantum magnets
have been studied mainly by means of the chain mean-field
approximation, where interchain spin fluctuations are ignored
completely.\cite{ScalapinoIP1975,Schulz1996}  Within the framework of
the chain mean-field approximation, a Q1D magnet is reduced to a
single chain in an effective external field.  For the latter system,
fortunately there exist a couple of exact solutions, otherwise one can
still use powerful analytic methods, such as bosonization, as well as
numerical simulations, such as exact diagonalization or density-matrix
renormalization group method, which are effective especially in one
dimension.

Naively, one may expect that the chain mean-field approximation
becomes more and more accurate not only qualitatively but also
quantitatively, as the interchain interactions become weak enough
compared to those along the chains.  The recent theoretical
study\cite{IrkhinK2000} as well as the sensitive Monte Carlo
simulations\cite{YasudaTHAKTT2005,Todo2006} on Q1D spin models, however,
have revealed that this is not the case; there remains systematic
error of the chain mean-field theory even in the weak interchain
coupling limit.  Instead, the critical temperature as a function of
the interchain coupling is well described by a chain mean-field
formula with a renormalized effective coordination number (or
effective interchain coupling\footnote{Note that both interpretations
  are equivalent, since the coordination number $z$ and the absolute
  magnitude of the interchain coupling $|J'|$ always appear as a
  product, $z|J'|$, in the case of the Weiss-type chain mean-field
  theory [see Eq.~(\ref{eqn:sce-weiss})].}).  Especially, in the Q1D
classical Ising models, it is proved that the renormalization factor
is exactly given by the critical transverse field of the quantum phase
transition of a two-dimensional quantum Ising model.\cite{Todo2006}
This analytic result demonstrates clearly that the weak interchain
coupling limit of the Q1D magnet is {\em not} the weak coupling limit,
but is still the strongly correlated regime.

The renormalization of effective interplane coupling is also observed
in weakly-coupled two-dimensional planes.\cite{YasudaTHAKTT2005} For
this quasi-two-dimensional (Q2D) system, a scaling theory was
developed,\cite{HastingsM2006} in which it is expected different
scaling behavior depending on whether the purely two-dimensional
system is in a quantum critical regime or in a renormalized classical
regime.  This prediction has been verified by a recent quantum Monte
Carlo simulation.\cite{YaoS2007}

Thus it has become evident by the recent theoretical and numerical
studies that we need a theory beyond the conventional chain mean-field
approximation to describe the critical temperature of the Q1D magnets
in the weak interchain coupling regime more accurately.  Even worse,
when the chain mean-field theory is applied to a system with quenched
disorder, the random average in each chain is taken {\rm before} the
thermal average on the whole lattice.  The adverse impact of
interchanging the average operations is further nontrivial.
It is naturally expected that some of these drawbacks of the chain
mean-field theory might be remedied by considering multiple chains,
instead of a single chain, and taking the interactions between those
chains into account in a proper way.  Along this line, Sandvik has
proposed a multichain mean-field theory and applied it successfully to
the problem of ground state staggered magnetization in a
two-dimensional Heisenberg antiferromagnet.\cite{Sandvik1999}

In the present paper, we propose a different type of chain mean-field
theory, i.e., the chain Bethe approximation.  Actually, it is well
known that there are two different formulations in traditional
mean-field theories for classical spin models, that is, the Weiss
approximation~\cite{Weiss1907} and the Bethe
approximation.\cite{Bethe1935,Peierls1936}  In the former, the
effective field is identified explicitly with the order parameter.  In
the latter, on the other hand, the effective field is determined
implicitly so that the local order parameter at the central spin and
that of spins on the cluster boundary coincide with each other.  In
the Bethe mean-field theory, therefore, spin fluctuations between
nearest-neighboring sites are taken into account even in the lowest
order approximation.  Furthermore, as the cluster size increases, the
critical temperature by the Bethe-type approximation converges to the
exact value more rapidly, being free from logarithmic corrections
which is observed in the Weiss-type theory.\cite{SuzukiK1986}  By
applying the idea of the Bethe-type effective-field theory, we
introduce an new chain mean-field theory for Q1D quantum magnets.

The present paper is organized as follows: In Sec.~2, after a brief
review of the conventional chain mean-field approximation, we
formulate our Bethe-type chain mean-field theory, the chain Bethe
approximation.  In Sec.~3, the chain Bethe approximation is applied to
the Q1D Heisenberg antiferromagnets, where we show that by using our
new approximation, the accuracy of the critical temperature is
significantly improved, compared with the conventional chain
mean-field theory.  In addition, we demonstrate that the new theory
can predict a {\it lower} critical temperature for the ferromagnetic
interchain coupling than in the antiferromagnetic case, which is also
confirmed by the QMC simulation.  In Sec.~4, we apply the chain Bethe
theory to a random quantum magnet, the site-diluted Q1D Heisenberg
antiferromagnet, where the existence of a finite critical impurity
concentration, above which the long-range order does not emerge even
in the zero-temperature limit, is predicted by using the chain Bethe
approximation.  The final section is for a summary and discussion.

\section{Chain Bethe Mean-field Theory}

\subsection{Conventional chain mean-field approximation}

The Hamiltonian of spin-1/2 Heisenberg model on a Q1D simple cubic
lattice is defined by
\begin{equation}
  \begin{split}
    \mathcal{H} &= J
    \sum_{i,j,k} \Vec{S}_{i,j,k} \cdot \Vec{S}_{i,j,k+1} + J' \sum_{i,j,k}
    \Vec{S}_{i,j,k} \cdot \Vec{S}_{i+1,j,k} \\ & + J' \sum_{i,j,k}
    \Vec{S}_{i,j,k} \cdot \Vec{S}_{i,j+1,k} - h \sum_{i,j,k} \phi^{i+j}
    (-1)^k S^z_{i,j,k},
  \end{split}
  \label{eqn:hamiltonian-1}
\end{equation}
where $\Vec{S}_{i,j,k} = (S^x_{i,j,k}, S^y_{i,j,k}, S^z_{i,j,k})$ is
an $S=1/2$ quantum spin operator at site $(i,j,k)$.  We take the
lattice $z$-axis as the chain direction.  The first term in
Eq.~(\ref{eqn:hamiltonian-1}) represents the intrachain interactions
(with coupling constant $J$), while the second and third terms denote
the interchain interactions ($J'$).  In the present paper we consider
only the case where the intrachain coupling $J$ is antiferromagnetic
($J>0$), whereas the interchain coupling $J'$ is either
antiferromagnetic ($J'> 0$) or ferromagnetic ($J'<0$).  Generalization
to the case with ferromagnetic intrachain coupling is also
straightforward.  The last term in Eq.~(\ref{eqn:hamiltonian-1})
represents an external magnetic field conjugate to the order
parameter, where the phase factor $\phi$ is defined as $\phi=-{\rm
  sign}(J')$.

\begin{figure}
 \resizebox{0.38\textwidth}{!}{\includegraphics{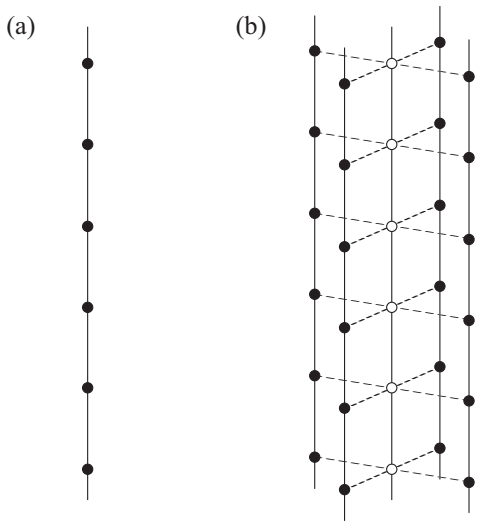}}
 \caption{Clusters for (a) chain Weiss and (b) chain Bethe
   approximations.  The solid and dashed lines denote the intrachain
   ($J$) and interchain ($J'$) interactions, respectively.  In the
   chain Weiss approximation, an alternating effective field with the
   same amplitude applies on all spins.  In the chain Bethe
   approximation, on the other hand, a staggered effective field
   applies only on the spins on the side chains (filled symbols).}
 \label{fig:clusters}
\end{figure}

In the conventional chain mean-field approximation, referred to as the
{\em chain Weiss theory} hereafter, spin fluctuations between the
chains are ignored and replaced by an effective field, i.e.,
\begin{equation}
 \begin{split}
   & J'\Vec{S}_{i,j,k} \cdot \Vec{S}_{i',j',k} \simeq J'
   \Vec{S}_{i,j,,k} \cdot \langle\Vec{S}_{i',j',k} \rangle \\
   & \qquad +
  J' \langle\Vec{S}_{i,j,k} \rangle \cdot \Vec{S}_{i',j',k}  + \mathrm{const.},
 \end{split}
\end{equation}
where $(i',j')=(i \pm 1,j)$ or $(i,j \pm 1)$.  As a result, the
original Hamiltonian~(\ref{eqn:hamiltonian-1}) is decoupled into a set
of independent chains~[Fig.~\ref{fig:clusters}(a)].  The effective
chain Hamiltonian for $(i,j)=(0,0)$ is written as
\begin{equation}
\begin{split}
\mathcal{H}_{\mathrm{c}} &= J \sum_k \Vec{S}_{k} \cdot
 \Vec{S}_{k+1} + J' \sum_k \Vec{S}_{k} \cdot \Vec{M}_{k}
 - h \sum_{k} (-1)^k S^z_{k}
 \end{split}
\label{eqn:hamiltonian-2}
\end{equation}
with $\Vec{S}_{k} \equiv \Vec{S}_{0,0,k}$ and
\begin{equation}
 \Vec{M}_{k} \equiv \langle\Vec{S}_{1,0,k}\rangle +
  \langle\Vec{S}_{-1,0,k}\rangle +
  \langle\Vec{S}_{0,1,k}\rangle +
  \langle\Vec{S}_{0,-1,k}\rangle.
  \label{eqn:def-M}
\end{equation}

In the low-temperature ordered phase, a finite magnetization appears
spontaneously even at $h=0$.  We assume the magnetization is along the
$z$-direction in the spin space, i.e., $\langle\Vec{S}_{i,j,k}\rangle
= (0, 0,\phi^{i+j}(-1)^k m(T))$.  The chain
Hamiltonian~(\ref{eqn:hamiltonian-2}) with $h=0$ is then reduced to
\begin{equation}
 \mathcal{H}_{\mathrm{c}} = J \sum_k \Vec{S}_{k} \cdot \Vec{S}_{k+1} -
  4 \, |J'| \, m(T) \sum_k (-1)^k S_k^z.
\label{eqn:hamiltonian-3}
\end{equation}
Since the magnitude of the spontaneous magnetization does not depend
on the position of the spin, the self-consistent condition
\begin{equation}
 m(T)=m_{\mathrm{c}}(T, 4 \, |J'| \, m(T))
  \label{eqn:self-consistent-equation}
\end{equation}
must be fulfilled, where
\begin{equation}
  m_{\mathrm{c}}(T,h) \equiv \frac{1}{L} \sum_k (-1)^k \langle S_k^z
  \rangle_\mathrm{c}
  \label{eqn:chain-mag}
\end{equation}
is the staggered magnetization density of genuinely one-dimensional
antiferromagnetic chain of length $L$.  The average $\langle \cdots
\rangle_\mathrm{c}$ in Eq.~(\ref{eqn:chain-mag}) means the expectation
value of the one-dimensional chain (i.e. Eq.~(\ref{eqn:hamiltonian-2})
with $J'=0$), while $\langle \cdots \rangle$ [e.g.\ in
Eqs.~(\ref{eqn:hamiltonian-1}) and (\ref{eqn:def-M})] denotes the
average with respect to the original (full-3D)
Hamiltonian~(\ref{eqn:hamiltonian-1}).

At high temperatures, on the other hand, no spontaneous magnetization
appears.  Under the presence of small external magnetic field,
however, a finite magnetization $\langle\Vec{S}_{i,j,k}\rangle =
(0,0,\phi^{i+j}(-1)^{k}m(T,h))$ is induced. In this case, the
effective chain Hamiltonian~(\ref{eqn:hamiltonian-3}) is modified as
\begin{equation}
  \begin{split}
    \mathcal{H}_{\mathrm{c}} = J \sum_k
    \Vec{S}_{k} \cdot \Vec{S}_{k+1} 
    - [h + 4 \, |J'| \, m(T,h)] \sum_k
    (-1)^k S_i^z.
  \end{split}
  \label{eqn:hamiltonian-4}
\end{equation}
Since $m(T,h) \ll 1$ for $h \ll 1$ in the disordered phase, one can
consider only the lowest order in $h$:
\begin{equation}
  m(T,h) \simeq [h+4 \, |J'| \, m(T,h)] \chi_{\mathrm{c}}(T),
\end{equation}
where $\chi_{\mathrm{c}}(T)$ is the zero-field staggered
susceptibility of one-dimensional antiferromagnetic chain.  Noticing
that $m(T,h)$ in the both sides can also be written as $h \chi(T)$ for
$h \ll 1$, we obtain the following mean-field expression for the
susceptibility:
\begin{equation}
  \chi(T) = \frac{\chi_{\mathrm{c}}(T)}{1-4|J'|\chi_{\mathrm{c}}(T)}.
  \label{eqn:cweiss-chi}
\end{equation}
In terms of the chain Weiss theory, the critical temperature is thus
given by the pole of the r.h.s.\ of Eq.~(\ref{eqn:cweiss-chi}).  For
generic Q1D lattices the self-consistent equation is written as
follows:
\begin{equation}
  1-z \, |J'| \, \chi_{\mathrm{c}}(T_\mathrm{c}) = 0,
  \label{eqn:sce-weiss}
\end{equation}
where $z$ is the coordination number of the lattice, i.e., the number
of nearest-neighbor chains.  Note that the self-consistent equation
depends not on the sign of $J'$, but only on its absolute magnitude.
In other words, the conventional chain Weiss theory does not
distinguish between the antiferromagnetic and ferromagnetic interchain
interactions.  This is one of the major drawbacks of the conventional
chain mean-field theory.  Another problem is that most of physical
quantities, such as the energy, specific heat, correlation functions,
etc, are the same as those of the genuine one-dimensional chain at
temperatures higher than $T_\mathrm{c}$, since the effective field is
proportional to the order parameter, which is zero at
$T>T_\mathrm{c}$.  We will see below that these disadvantages of the
chain Weiss approximation are solved in the chain Bethe mean-field
theory.

\subsection{Bethe-type mean-field theory}

In the chain Weiss approximation, the effect of interchain interaction
was replaced by an effective field.  The effective interchain
interaction is directly related with the order parameter, and
interchain spin fluctuations are thus ignored completely.  The
approximation can be improved by taking interchain spin fluctuations
into account systematically.  In the chain Bethe approximation
introduced in this section, the interaction between the
nearest-neighbor chains are taken into account exactly, and those
around the multichain cluster are treated as an effective field.  In
the Bethe approximation,\cite{Bethe1935,Peierls1936} the effective
field is determined so that the magnetization of the central spin and
that on the cluster boundary coincide with each other.  On the
contrary to the Weiss theory, the order parameter is not given
explicitly by the effective field, but is its implicit function.

Let us consider the chain Bethe approximation for the simple cubic
lattice.  In this case we prepare a cluster of five
chains~[Fig.~\ref{fig:clusters}(b)], where an effective field is
applied only to the spins on the side chains (black circles).  The
effective Hamiltonian is written as
\begin{equation}
  \begin{split}
    \mathcal{H}_\mathrm{\Omega} &=  J \sum_{\alpha=0}^4 \sum_k \Vec{S}_{\alpha,k} \cdot \Vec{S}_{\alpha,k+1}
  + J' \sum_{\alpha=1}^4 \sum_k \Vec{S}_{0,k} \cdot \Vec{S}_{\alpha,k} \\
  &- h \sum_k (-1)^k S_{0, k}^{z}
  - (h + h_\mathrm{eff}) \sum_{\alpha=1}^4 \sum_k \phi (-1)^k S_{\alpha, k}^{z},
  \end{split}
\end{equation}
where $\alpha=0$ denotes the central chain [$(i,j)=(0,0)$] and
$\alpha=1,\cdots,4$ denotes the side chains [$(i,j)=(\pm 1,0)$ or
$(0,\pm 1)$].

In the chain Bethe approximation, we impose the condition that the
absolute value of the local magnetization does not depend on its
position:
\begin{equation}
  \begin{split}
    (-1)^k \langle \Vec{S}_{0,k} \rangle_\Omega = 
    \phi (-1)^{k'} \langle \Vec{S}_{\alpha,k'} \rangle_\Omega
  \end{split}
\end{equation}
for any $k$, $k'$, and  $\alpha=1,\cdots,4$, or equivalently
\begin{equation}
  m_{\Omega,0}(T, h, h_\textrm{eff}) = \frac{\phi}{4} \sum_{\alpha=1}^4 m_{\Omega,\alpha}(T, h, h_\textrm{eff}),
  \label{eqn:cbethe-sce}
\end{equation}
where $m_{\Omega,\alpha}(T, h, h_\textrm{eff})$ is the staggered
magnetization density of the $\alpha$-th chain:
\begin{equation}
  m_{\Omega,\alpha}(T, h, h_\textrm{eff}) = \frac{1}{L} \sum_k (-1)^k \langle \Vec{S}_{\alpha,k} \rangle_\Omega.
\end{equation}
Here $\langle \cdots \rangle_\Omega$ denotes the expectation value of
the chain Bethe cluster and $L$ the number of spins in the chain direction.

\begin{figure}
 \resizebox{0.48\textwidth}{!}{\includegraphics{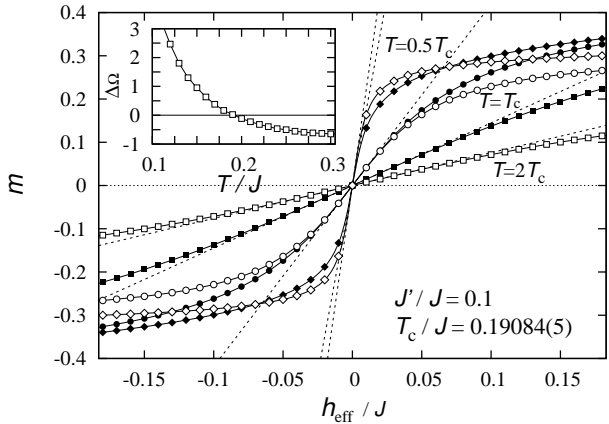}}
 \caption{$h_{\rm eff}$-dependence of the local magnetization of the
   center spins (open symbols) and the boundary spins (filled symbols)
   of $L=64$ chain Bethe cluster with $J'/J=0.1$.  At temperatures higher
   than the critical temperature (squares) two curves intersect only
   at $h_{\rm eff}=0$, while at lower temperature (diamonds)
   nontrivial solutions corresponding to the symmetry-broken phases
   appear.  The dashed lines denote the tangent of each curve at
   $h_{\rm eff}=0$.  In the inset, temperature dependence of
   $\Delta_\Omega$ [Eq.(\ref{eqn:cbethe-sce2})] is also presented.}
 \label{fig:cbethe-0100}
\end{figure}

In Fig.~\ref{fig:cbethe-0100}, the $h_\textrm{eff}$ dependence of the
both sides in Eq.~(\ref{eqn:cbethe-sce}) is demonstrated for $J' =
0.1J$ and $h=0$, where the magnetization is calculated by means of
the quantum Monte Carlo (QMC) method for an $L=64$ chain Bethe cluster (see
Sec.~\ref{sec:qmc} for simulation details).  It is clearly seen that
Eq.~(\ref{eqn:cbethe-sce}) has only one trivial solution,
$h_\textrm{eff} = 0$, at high temperatures.  On the other hand, at low
temperatures two more nontrivial solutions ($h_\textrm{eff} \ne 0$)
appear, which correspond to the the symmetry-broken phase.

The critical temperature in the framework of the chain Bethe
approximation is defined as the point where the three solutions at low
temperatures get degenerated with each other
(Fig.~\ref{fig:cbethe-0100}).  In practice, we set $h=0$ in
Eq.~(\ref{eqn:cbethe-sce}) and expand the both sides in terms of
$h_\textrm{eff}$.  The self-consistent equation for the critical
temperature is then written as
\begin{equation}
  \Delta_\Omega(T_\textrm{c}) \equiv J \Big[ \chi_{\Omega,0}(T_\textrm{c}) -
  \frac{\phi}{4} \sum_{\alpha=1}^4 \chi_{\Omega,\alpha}(T_\textrm{c}) \Big] = 0
  \label{eqn:cbethe-sce2}
\end{equation}
with the boundary-field susceptibilities
\begin{equation}
  \begin{split}
    \chi_{\Omega,\alpha}(T) &= \left. \frac{\partial m_{\Omega,\alpha}(T, h, h_\textrm{eff})}{\partial h_\textrm{eff}} \right|_{h=0, h_\textrm{eff}=0} \\
    &= \frac{\beta}{L} \sum_{\alpha'=1}^4 \sum_{k,k'} \left\langle
      S_{\alpha,k}^z;S_{\alpha',k'}^z \right\rangle_\Omega ,
  \end{split}
  \label{eqn:locsus}
\end{equation}
where $\left\langle A;B \right\rangle_\Omega$ denotes the canonical
correlation:
\begin{equation}
  \left\langle A;B \right\rangle_\Omega = 
  \frac{1}{\beta} \frac{\displaystyle \textrm{Tr} \! \int_0^\beta \!\! A \, e^{-\tau {\cal H}_\Omega} B \, e^{-(\beta -\tau){\cal H}_\Omega} \, d\tau}{\displaystyle \textrm{Tr} \, e^{-\beta {\cal H}_\Omega}}
\end{equation}
of two operators $A$ and $B$.  In the inset of
Fig.~\ref{fig:cbethe-0100}, we show the temperature dependence of
$\Delta_\Omega(T)$.  As the temperature increases, $\Delta_\Omega(T)$
decreases monotonically.  The critical temperature, in terms of the
chain Bethe approximation, is given as the zero of $\Delta_\Omega(T)$.
For $J'/J=0.1$, we obtained $T_\mathrm{c}/J = 0.19084(5)$.

\section{Critical temperature of Q1D quantum Heisenberg models}

\subsection{Numerical method}
\label{sec:qmc}

In this section, we discuss the interchain coupling dependence of the
critical temperature for the $S=1/2$ Heisenberg model on a Q1D simple
cubic lattice [Eq.~(\ref{eqn:hamiltonian-1})] in terms of the chain
Bethe approximation.  Since the boundary-field susceptibility of the chain
Bethe cluster [Eq.~(\ref{eqn:locsus})] can not be evaluated
analytically, one needs to introduce some reasonable approximation or
some numerical method.  In the present paper, we adopt the
continuous-time loop cluster QMC algorithm, which is one of the most
effective methods for simulating unfrustrated quantum spin
systems.\cite{Evertz2003,TodoK2001}  It is a variant of the world-line QMC
method, based on the Suzuki-Trotter path-integral expansion.  The
continuous-time loop algorithm, however, works directly in the
imaginary-time continuum, and thus it is completely free from the time
discretization error.  Furthermore, the correlation between successive
spin configurations on the Markov chain is greatly reduced, often by
several orders of magnitude.  This is manifested by the fact that
clusters of spins, called loops, whose linear size corresponds
directly to the length scale of relevant spin fluctuations, are
flipped at once in the loop algorithm.

The boundary-field susceptibilities~(\ref{eqn:locsus}) are calculated
by means of the improved estimator.\cite{Evertz2003}  The largest
system size in the chain direction is $L=64$ for $|J'|/J \ge 0.05$.
For smaller $|J'|$'s, longer systems (e.g., $L=512$ for $|J'|/J =
0.01$) are needed to obtain the susceptibilities in the thermodynamic
limit $L \rightarrow \infty$, since the critical temperature decreases
as $|J'|$ does.  Periodic boundary conditions are imposed along the
chains.  We interpolate the QMC results at various temperatures by a
polynomial, and estimate the zero of $\Delta_\Omega(T)$, which gives
the critical temperature.  (See e.g., the inset of
Fig.~\ref{fig:cbethe-0100}.)

\begin{figure}
 \resizebox{0.48\textwidth}{!}{\includegraphics{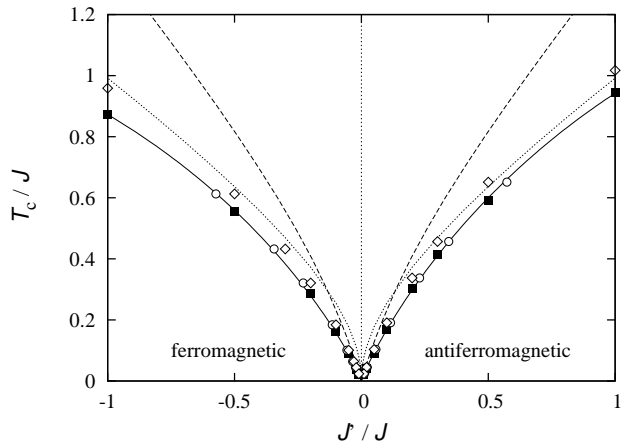}}
 \caption{$J'$-dependence of the critical temperature calculated by
   chain Bethe theory (diamonds), chain Weiss theory (dashed lines),
   Tyablikov approximation (dotted lines), and full-3D QMC (filled
   squares), for antiferromagnetic (right) and ferromagnetic (left)
   interchain coupling.  The QMC results for the antiferromagnetic
   interchain coupling are taken from
   Ref.~\onlinecite{YasudaTHAKTT2005}.  The chain Weiss and Tyablikov
   approximations both produce the identical results for the
   antiferromagnetic and ferromagnetic cases.  The solid lines are the
   guide to the eye.  The error bar of each data is much smaller than
   the symbol size.  The results of modified chain Bethe approximation
   with $J'_{\rm eff}=0.872J'$ (see the text) are also presented by
   open circles.}
 \label{fig:pure-tc}
\end{figure}

\subsection{Comparison of chain Bethe theory with other
  methods}

The accuracy of the present chain Bethe approximation is checked by
comparing with the results of QMC simulations of full-3D systems,
i.e., simple cubic lattice of $L_x \times L_y \times L_z$ sites with
periodic boundary conditions along all the lattice axes.  For the
antiferromagnetic interchain interactions, $J'/J>0$, we take the
full-3D QMC results from Ref.~\onlinecite{YasudaTHAKTT2005}.  For
$J'/J<0$, the critical temperature is estimated from the crossing
point of the Binder cumulant
\begin{align}
  Q = \frac{\langle m^2 \rangle^2}{\langle m^4 \rangle}
  \label{eqn:binder}
\end{align}
of the generalized magnetization density
\begin{align}
  m = \frac{1}{L_x L_y L_z} \sum_{i,j,k} (-1)^k S_{i,j,k}^z
\end{align}
for different system sizes.  The largest system size we simulate is
$(L_x,L_y,L_z) = (40,40,40)$ and $(12,12,264)$ for $|J'|/J=1$ and
$0.01$, respectively.  The results of finite-size scaling are
summarized in Table~\ref{tbl:ferro-tc}.

\begin{table}
  \caption{$J'$-dependence of the critical temperature
    for antiferromagnetic ($J'>0$) and ferromagnetic ($J'<0$)
    interchain interactions obtained by the full-3D QMC calculation.
    The results for $J'>0$ are from the previous QMC
    study.\cite{YasudaTHAKTT2005}
    The figure in the parenthesis denotes the error
    in the last digit.}
  \label{tbl:ferro-tc}
  \begin{tabular}{lllll}
    \hline
    \hline
    \multicolumn{1}{c}{$|J'|/J$} & $\qquad$ & \multicolumn{3}{c}{$T_{\rm c}/J$}
    \\
    \multicolumn{1}{c}{} & $\qquad$ & \multicolumn{1}{c}{$J'>0$} & $\qquad$ &
    \multicolumn{1}{c}{$J'<0$} \\
    \hline
    \ \ 1     & & 0.94416(9)  & & 0.87330(2) \ \\
    \ \ 0.5   & & 0.59248(6)  & & 0.55419(3) \ \\
    \ \ 0.3   & & 0.4151(2)   & & \multicolumn{1}{c}{---} \ \\
    \ \ 0.2   & & 0.30202(5)  & & 0.28685(2) \ \\
    \ \ 0.1   & & 0.16917(2)  & & 0.16295(2) \ \\
    \ \ 0.05  & & 0.09129(4)  & & 0.08900(3) \ \\
    \ \ 0.02  & & 0.039432(7) & & 0.03899(2) \ \\
    \ \ 0.01  & & 0.020763(8) & & 0.02080(2) \ \\
    \hline
    \hline
  \end{tabular}
\end{table}

In Fig.~\ref{fig:pure-tc}, we show the $J'$-dependence of the critical
temperature, calculated in terms of the chain Bethe approximation,
both for antiferromagnetic ($J' > 0$) and ferromagnetic ($J' < 0$)
interchain interactions, together with the full-3D QMC results.  The
intrachain interaction is antiferromagnetic ($J > 0$) in either case.
In Fig.~\ref{fig:pure-tc}, we also present the results of the chain
Weiss approximation and the Green's function method with Tyablikov
decoupling.\cite{Tyablikov1959,OguchiH1963}  As for the chain Weiss
theory, we use the susceptibility of genuinely one-dimensional
antiferromagnetic chain calculated by the QMC method for the
antiferromagnetic chain of 256 spins.

In both of antiferromagnetic and ferromagnetic interchain coupling
cases, the critical temperature decreases monotonically as $|J'|/J$
does.  For $|J'| \approx J$, it is observed that the chain Weiss
theory overestimates the critical temperature greatly.  This is is
because this approximation ignores the interchain spin fluctuations
completely.  On the other hand, for $|J'| \ll J$ the Tyablikov
approximation gets worse.  Indeed, it predicts the critical
temperature proportional to the square root of $|J'|/J$, which does
not agree with the correct asymptotic behavior, $T_{\rm c} \sim
|J'|/J$ (with some logarithmic
corrections).\cite{ScalapinoIP1975,Schulz1996}  It should be noted
that the chain Weiss theory and the Tyablikov approximation both give
the identical critical temperature dependence regardless of the sign
of the interchain coupling.

The results of the chain Bethe approximation are fairy well in the
whole region (Fig.~\ref{fig:pure-tc}).  The relative errors from the full-3D QMC values are
about 7--10\% at $|J'|/J=1$, which should be compared with the
conventional chain Weiss results (48--60\%).  Surprisingly, the
present chain Bethe approximation predicts a {\em lower} critical
temperatures for the ferromagnetic interchain coupling than the
antiferromagnetic case, e.g., $T_{\rm c}/J = 1.017$ and 0.958 for
$J'/J=1$ and -1, respectively.  This result is seemingly
counter-intuitive, since quantum fluctuations, which is expected to
suppress the classical ordering, are generally much stronger in the
antiferromagnetic cases.  However, it is not an artifact by our
approximation.  Indeed, a lower critical temperature for the
ferromagnetic interchain coupling is also confirmed by our full-3D QMC
calculation (Table~\ref{tbl:ferro-tc}).  Thus, the chain Bethe
approximation predicts not only quantitatively more accurate critical
temperatures, but also its nontrivial dependence on the sign of the
interchain coupling.

\subsection{Effective interchain interaction}

\begin{figure}
 \resizebox{0.48\textwidth}{!}{\includegraphics{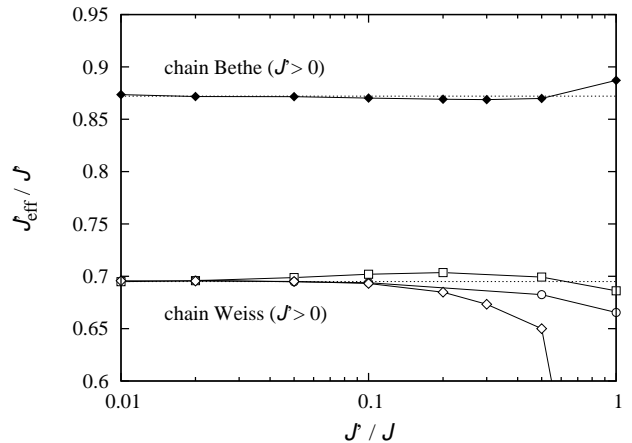}}
 \caption{$J'$-dependence of the renormalization factor $J'_{\rm
     eff}/J'$ for chain Bethe (filled diamonds) and chain Weiss ($S=1/2$
   open diamonds, $S=3/2$ circles, $S=\infty$ squares) approximations.
   The horizontal dashed lines denote their limiting values (0.872 and
   0.695, respectively) for $J'/J \rightarrow 0$.}
 \label{fig:effective-J'}
\end{figure}

In order to evaluate the accuracy of the present theory for small
$|J'|/J$ in a more systematic way, next we discuss the effective
interchain coupling, $J'_{\rm eff}(J')$.  This quantity, firstly
introduced in Ref.~\onlinecite{YasudaTHAKTT2005}, is defined as the
coupling constant which predicts the true critical temperature if it
is used in the self-consistent equation instead of the original $J'$.
For the chain Weiss approximation, $J'_{\rm eff}$ is explicitly
obtained from Eq.~(\ref{eqn:sce-weiss}) as
\begin{equation}
  J'_{\rm eff}(J') = \frac{{\rm sign}(J')}{z \chi_{\rm c}(T_{\rm c}(J'))} ,
\end{equation}
where we use the full-3D QMC results for $T_{\rm c}(J')$
(Table.~\ref{tbl:ferro-tc}), which is considered to be exact within
the error bar.  On the other hand, for the chain Bethe approximation,
we obtained $J'_\textrm{eff}$ by solving Eq.~(\ref{eqn:cbethe-sce2})
numerically.  In fig.~\ref{fig:effective-J'}, we plot the
$J'$-dependence of the effective interchain coupling for chain Bethe
approximation ($J'>0$ and $S=1/2$) together with those of the chain
Weiss approximation ($J'>0$ and $S=1/2$, 3/2, and
$\infty$).\cite{YasudaTHAKTT2005}  In both cases, $J'_\textrm{eff}/J'$
converges to a finite value for $J'/J \ll 1$.  The limiting value is
0.872 and 0.695 for the chain Bethe and Weiss approximations,
respectively.  The larger (or closer to unity) value of
$J'_\textrm{eff}/J'$ in the former supports that it is indeed a better
approximation compared with the latter.  Interestingly, the chain
Bethe approximation has the largest $J'_\textrm{eff}/J'$ at $J'/J=1$,
in other words, one may say that it becomes the most accurate in the
isotropic limit from the viewpoint of the renormalized factor of
interchain coupling.

On should note that, in the chain Bethe approximation, not only the
renormalization factor $J'_\textrm{eff}/J'$ is improved very much, but
also it converges to its limiting value quite rapidly (already
converged at $J'/J = 0.5$).  This result suggests the critical
temperature might be well described for any values of $J'/J$ by the
chain Bethe approximation using the renormalized interchain coupling
constant, $0.872 J'$, instead of the bare interchain coupling, $J'$
(modified chain Bethe approximation).  In Fig.~\ref{fig:pure-tc}, we
also plot the result of the modified chain Bethe theory by open
circles, which satisfactorily agrees with the true critical
temperature in the whole range of $|J'|/J$.  Note that in this plot we
use the same renormalization factor, 0.872, both for the
antiferromagnetic and ferromagnetic interchain coupling cases.

\section{Site-diluted Q1D Heisenberg antiferromagnet}

In the previous section, we see that the critical temperature of Q1D
system is quantitatively improved greatly by the chain Bethe
approximation.  In this section, we discuss a more nontrivial example,
the site-diluted Q1D Heisenberg antiferromagnet, where the chain Weiss
approximation fails even qualitatively.

The Hamiltonian of the site-diluted Heisenberg antiferromagnet is
defined as follows:
\begin{equation}
  \begin{split}
    \mathcal{H}
    &= J \sum_{i,j,k} \epsilon_{i,j,k} \epsilon_{i,j,k+1} \Vec{S}_{i,j,k} \cdot \Vec{S}_{i,j,k+1} \\
    & + J' \sum_{i,j,k} \epsilon_{i,j,k} \epsilon_{i+1,j,k} \Vec{S}_{i,j,k} \cdot \Vec{S}_{i+1,j,k} \\
    & + J' \sum_{i,j,k} \epsilon_{i,j,k} \epsilon_{i,j+1,k} \Vec{S}_{i,j,k} \cdot \Vec{S}_{i,j+1,k},
  \end{split}
  \label{eqn:hamiltonian-dilute}
\end{equation}
where $\{ \epsilon_{i,j,k} \}$ are the quenched dilution factors.
They take either 1 (occupied) or 0 (vacant) independently with
probability $(1-x)$ and $x$, respectively, with $x$ ($0 \le x \le 1$) being the
concentration of vacancies, or nonmagnetic impurities.

The ground state of the classical site-diluted spin model is
equivalent to the site-percolation problem.\cite{StaufferA1994}  The
system undergoes a second-order phase transition at the percolation
threshold $x_{\rm p}$, above which there exist no infinite-size
clusters.  For the simple cubic lattice, the percolation threshold is
determined as
\begin{align}
  x_{\rm p} = 0.6883923(4)
\end{align}
by the most recent simulation.\cite{DengB2005}  In the quantum spin
cases, whether a long-range order exists or not near the percolation
threshold is a nontrivial problem due to the presence of quantum
fluctuations.  However, it has been established by the extensive QMC
simulation that the staggered magnetization persists up to the
percolation threshold on the two-dimensional square
lattice.\cite{KatoTHKMT2000}  We expect this is also the case for the
present anisotropic simple cubic lattice.

\begin{figure}
 \resizebox{0.48\textwidth}{!}{\includegraphics{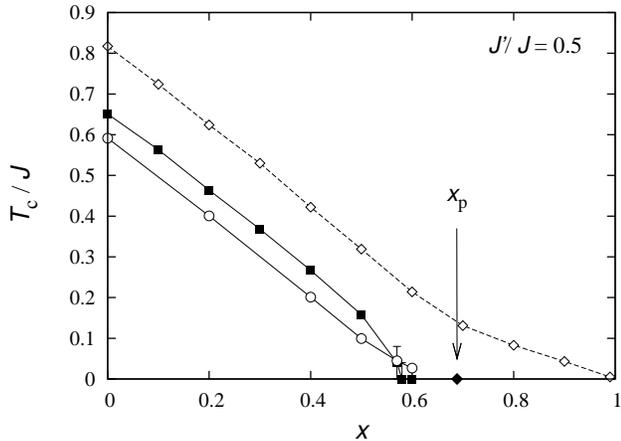}}
 \caption{$x$-dependence of the N\'eel temperature of the $S=1/2$
   diluted Heisenberg antiferromagnet with $J'/J=0.5$ obtained by the
   full-3D QMC calculation (open circles), the chain Weiss
   approximation (open diamonds), and the chain Bethe approximation
   (solid squares).  The percolation threshold of the simple cubic
   lattice, $x_{\rm p} \simeq 0.688$, is indicated by the arrow.}
 \label{fig:random-tc}
\end{figure}

In Fig.~\ref{fig:random-tc}, we show the $x$-dependence of the
critical temperature for $J'/J=0.5$ (antiferromagnetic interchain
interaction) obtained by the chain Weiss and chain Bethe
approximations together with the results of the full-3D QMC
simulation.  The largest system size used in the full-3D QMC
simulation is $(L_x,L_y,L_z)=(16,16,64)$.  The N\'eel temperature is
estimated from the crossing point of the Binder
cumulant~(\ref{eqn:binder}) of the staggered magnetization for
different system sizes.  From the full-3D QMC calculations, we thus
confirm that the N\'eel temperature remains finite at least up to
$x=0.6$.  For larger impurity concentration, it is not very easy to
estimate the critical temperature with satisfactory accuracy in the
present scale of simulation.  The staggered susceptibility of
one-dimensional chain used in the chain Weiss approximation is also
evaluated by means of the QMC method.  We simulate chains with spins
$L=128$ and 256 for various impurity concentrations and confirm that
there is no significant systematic difference in the results for those
two system sizes.  The N\'eel temperature is then estimated by solving
the self-consistent equation~(\ref{eqn:sce-weiss}) numerically.  Since
the staggered susceptibility of one-dimensional chain diverges
monotonically as the temperature decreases irrespective of impurity
concentration, the self-consistent equation always has a solution (see
the discussion below).

\begin{figure}
 \resizebox{0.48\textwidth}{!}{\includegraphics{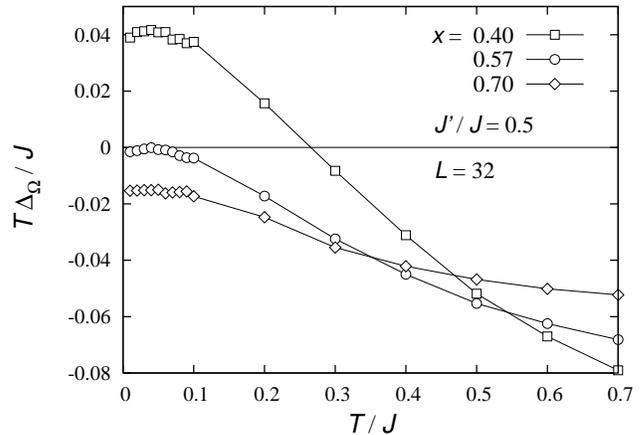}}
 \caption{Temperature dependence of $T \Delta_\Omega / J$
   [Eq.~(\ref{eqn:cbethe-sce2})] of the $S=1/2$ diluted Heisenberg
   antiferromagnet with $J'/J=0.5$ at $x=0.40$ (squares), 0.57
   (circles), and 0.70 (diamonds) obtained by the QMC method for
   the five-chain cluster of length $L=32$.  The error bars are smaller
   than the symbol size.}
 \label{fig:random-sus}
\end{figure}

For the chain Bethe approximation, we evaluate $\Delta_\Omega(T)$
[Eq.~(\ref{eqn:cbethe-sce2})] by calculating the boundary-field
susceptibilities of five-chain clusters with $L=32$ and 64 by means of the
QMC method.  We observe any significant differences between the $L=32$
and 64 results for $0 \le x \le 0.6$ in the temperature range we
simulate ($T/J \ge 0.01$).  We find that the function
$\Delta_\Omega(T)$ tends to positively or negatively diverge depending
on the impurity concentration.  To see the tendency at low
temperatures more clearly, we plot $T \Delta_\Omega(T) / J$, instead of
$\Delta_\Omega(T)$ itself, as a function of temperature in
Fig.~\ref{fig:random-sus}.  At low enough temperatures the quantity
tends to converge to a finite value, which gives the coefficient of
Curie-like behavior of $\Delta_\Omega(T)$.  It is clearly seen that
for $x < 0.57$, the coefficient is positive, and thus the
self-consistent equation~(\ref{eqn:cbethe-sce2}) has a solution, while
the coefficient is negative and $\Delta_\Omega(T)$ has no zero
for $x > 0.57$.

In all the cases, the N\'eel temperature decreases monotonically as $x$
increases as shown in Fig.~\ref{fig:random-tc}.  However, we emphasize
that the result of the chain Weiss approximation is qualitatively
different from the others for large $x$; it predicts nonvanishing
N\'eel temperature for any $x<1$, though the others has a finite
critical concentration of impurities ($x_{\rm c} \simeq x_{\rm p}$ and
$x_{\rm c} \simeq 0.57$ for QMC and the chain Bethe approximation,
respectively).  Since no long-range order can persist for $x > x_{\rm
  p}$, the result of the chain Weiss theory in this regime is
unphysical at all.

Indeed, the asymptotic behavior of $T_{\rm c}$ near $x = 1$ can be
discussed more precisely as follows: In the chain Weiss theory, the
staggered susceptibility of the purely one-dimensional chain appears
in the self-consistent equation~(\ref{eqn:sce-weiss}).  The
percolation threshold of a single chain is unity, i.e., the chain is
decoupled into a set of finite-length segments immediately by an
infinitesimal impurity density.  Thus the staggered susceptibility can
be expressed as a weighted average of contributions from finite-length
segments:
\begin{align}
  \chi_{\rm c} (T) &= \sum_{\ell = 1} p_\ell \chi_{\ell}(T),
\end{align}
where $\chi_{\ell}(T)$ is the staggered susceptibility of a finite
segment of length $\ell$ and $p_\ell \equiv (1-x)^{\ell} x^2$ the
average number of segment of length $\ell$ per site.  For $(1-x) \ll
1$, only single-site clusters ($\ell = 1$) contribute to the
susceptibility:
\begin{align}
  \chi_{\rm c} (T) &= (1-x) \frac{1}{4T} + {\cal O}((1-x)^2).
\end{align}
By solving the self-consistent equation~(\ref{eqn:sce-weiss}), the
critical temperature is then obtained as
\begin{align}
  T_{\rm c} &= (1-x)\frac{zJ'}{4} + {\cal O}((1-x)^2).
  \label{eqn:weiss-asympt}
\end{align}
This expression gives the exact asymptotic behavior of $T_{\rm c}$ of
the chain Weiss theory in the vicinity of $x=1$.  For $J'/J=0.5$ and
$x=0.9$, Eq.~(\ref{eqn:weiss-asympt}) gives $T_{\rm c}/J = 0.05$,
which agrees fairly well with the result of the chain Weiss
approximation, $T_{\rm c}/J = 0.043$.

A similar discussion applies also to the chain Bethe approximation.
For $x \approx 1$, only single-site clusters contribute to the
susceptibility.  Since in the chain Bethe approximation the effective
field is applied only on the side chains, a single-site cluster on the
central chain does not feel the effective field, and thus the
boundary-field susceptibility vanishes in the lowest order:
\begin{align}
  \chi_{\Omega,0} &= {\cal O}((1-x)^2).
\end{align}
On the other hand, the susceptibility of boundary spins is given by the same
expression as in the chain Weiss approximation:
\begin{align}
  \chi_{\Omega,\alpha} &=(1-x)\frac{1}{4T} + {\cal O}((1-x)^2)
  \ \ \text{for $\alpha = 1,\cdots,4$.}
\end{align}
If these two expressions are substituted into the self-consistent
equation~(\ref{eqn:cbethe-sce2}), one immediately finds that it has no
solution for $0 \le x \le 1$.  If one further considers contribution
from dimers (i.e., clusters consist of two sites) the boundary-field
susceptibilities are calculated as
\begin{align}
  \chi_{\Omega,0} &= (1-x)^2 \frac{1}{2J'} + {\cal O}((1-x)^3) \\
  \begin{split}
    \chi_{\Omega,\alpha} &= (1-x)\frac{1}{4T} + (1-x)^2 \frac{1}{2J'} + 2 (1-x)^2 \frac{1}{J} \\ & + {\cal O}((1-x)^3) \ \ \text{for $\alpha = 1,\cdots,4$.}
  \end{split}
\end{align}
Again the self-consistent equation has no solution for $J'/J \ge 3/4$.
On the other hand, for a smaller $J'$ ($J'/J < 3/4$), there exists a
solution:
\begin{align}
  T_{\rm c} \approx \frac{1}{2(1-x)(\frac{3}{4} - \frac{J'}{J})}.
\end{align}
However, this solution is unphysical, since it diverges as $x
\rightarrow 1$.  We infer that even how one takes higher-order
contribution from large clusters into account, there exists no
physical solution of the self-consistent equation.  This implies that
the chain Bethe approximation has a finite critical threshold $x_c <
1$, above which no long-range order appears at finite temperatures.

Before closing this section, we briefly mention the initial reduction
rate of the critical temperature:
\begin{align}
  R = -\frac{d \log T_{\rm c}(x)}{dx} \Big|_{x=0}.
\end{align}
From the present QMC results, this quantity is estimated as $R = 1.61$
for $J'/J = 0.5$, which is significantly larger than that of the
isotropic cubic lattice ($R=1.22$ and 1.36 from
renormalization-group~\cite{Stinchcombe1979} and
series~\cite{RushbrookeMSP1972} studies, respectively).  Accordingly,
the critical temperature is a convex function of the impurity
concentration, which is in a sharp contrast to the linear behavior
observed in the isotropic cubic lattice.\cite{Stinchcombe1979}  Such
a large initial reduction rate is also observed experimentally in
quasi-two-dimensional Heisenberg antiferromagnet.\cite{CheongCRB1991}
The enhancement in the initial reduction rate and the convexity might
be attributed to the spatial anisotropy of the lattice.

\section{Summary}

In this paper, we proposed a novel chain Bethe theory for Q1D quantum
magnets.  In the present approximation, the self-consistent equation
is written in terms of the boundary-field magnetic susceptibilities of
a multichain cluster instead of a single chain.  Not only the
correlations along the chains, but also those between the
nearest-neighboring chains are taken into account exactly.  As a
result, the accuracy of the critical temperature of the Q1D Heisenberg
models is improved greatly compared with the conventional chain Weiss
theory.  It is also demonstrated that our new approximation can
predict nontrivial dependence of critical temperature on the sign of
interchain coupling as well as on the impurity concentration in
randomly diluted Q1D Heisenberg magnets.  The conventional chain Weiss
approximation takes the random average in each chain before the
thermal average on the whole lattice, whereas the present theory can
take fluctuations due to the randomness between the neighboring chains
effectively.  This difference in the order of thermal and random
averaging has a great impact especially in the system with strong
quenched disorder.

In the present study, we restricted ourselves to the nearest-neighbor
spin models on the simple cubic lattice.  This is because unbiased
high-precision full-3D data, by which the accuracy of the new theory
has been checked quantitatively, are available only for such
unfrustrated models.  It should be emphasized that, however, with the
help of other numerical methods specialized to one-dimensional
systems, such as the exact diagonalization and the density-matrix
renormalization group method, the present chain Bethe theory itself
can be applied straightforwardly to spin models with strong
frustration or even to fermionic models.  In such models, effects of
correlations between neighboring chains are much more important, and
thus the improved chain mean-field approach formulated in the present
paper could be an essential tool to investigate exotic phase
transitions as well as anomalous low-energy properties.

\section*{Acknowledgement}

Part of the simulations in the present paper has been done by using
the facility of the Supercomputer Center, Institute for Solid State
Physics, University of Tokyo.  The simulation code is developed based
on the ALPS/looper library.\cite{LOOPERweb,TodoK2001,ALPS2007}  One
of the authors (S.T.) acknowledges support by Grant-in-Aid for
Scientific Research Program (No.~18540369) from JSPS, and also by the
Grand Challenge to Next-Generation Integrated Nanoscience, Development
and Application of Advanced High-Performance Supercomputer Project
from MEXT, Japan.

\bibliography{main}

\begin{thebibliography}{27}
\expandafter\ifx\csname natexlab\endcsname\relax\def\natexlab#1{#1}\fi
\expandafter\ifx\csname bibnamefont\endcsname\relax
  \def\bibnamefont#1{#1}\fi
\expandafter\ifx\csname bibfnamefont\endcsname\relax
  \def\bibfnamefont#1{#1}\fi
\expandafter\ifx\csname citenamefont\endcsname\relax
  \def\citenamefont#1{#1}\fi
\expandafter\ifx\csname url\endcsname\relax
  \def\url#1{\texttt{#1}}\fi
\expandafter\ifx\csname urlprefix\endcsname\relax\def\urlprefix{URL }\fi
\providecommand{\bibinfo}[2]{#2}
\providecommand{\eprint}[2][]{\url{#2}}

\bibitem[{\citenamefont{Mermin and Wagner}(1966)}]{MerminW1966}
\bibinfo{author}{\bibfnamefont{N.~D.} \bibnamefont{Mermin}} \bibnamefont{and}
  \bibinfo{author}{\bibfnamefont{H.}~\bibnamefont{Wagner}},
  \bibinfo{journal}{Phys. Rev. Lett.} \textbf{\bibinfo{volume}{17}},
  \bibinfo{pages}{1133} (\bibinfo{year}{1966}).

\bibitem[{\citenamefont{Bethe}(1931)}]{Bethe1931}
\bibinfo{author}{\bibfnamefont{H.}~\bibnamefont{Bethe}}, \bibinfo{journal}{Z.
  Physik} \textbf{\bibinfo{volume}{71}}, \bibinfo{pages}{205}
  (\bibinfo{year}{1931}).

\bibitem[{\citenamefont{Schollw\"ock et~al.}(2004)\citenamefont{Schollw\"ock,
  Richter, Farnell, and Bishop}}]{SchollwoeckRFB2004}
\bibinfo{editor}{\bibfnamefont{U.}~\bibnamefont{Schollw\"ock}},
  \bibinfo{editor}{\bibfnamefont{J.}~\bibnamefont{Richter}},
  \bibinfo{editor}{\bibfnamefont{D.~J.~J.} \bibnamefont{Farnell}},
  \bibnamefont{and} \bibinfo{editor}{\bibfnamefont{R.~F.}
  \bibnamefont{Bishop}}, eds., \emph{\bibinfo{title}{Quantum Magnetism, Lecture
  Notes in Physics 645}} (\bibinfo{publisher}{Springer Verlag},
  \bibinfo{address}{Berlin}, \bibinfo{year}{2004}).

\bibitem[{\citenamefont{Scalapino et~al.}(1975)\citenamefont{Scalapino, Imry,
  and Pincus}}]{ScalapinoIP1975}
\bibinfo{author}{\bibfnamefont{D.~J.} \bibnamefont{Scalapino}},
  \bibinfo{author}{\bibfnamefont{Y.}~\bibnamefont{Imry}}, \bibnamefont{and}
  \bibinfo{author}{\bibfnamefont{P.}~\bibnamefont{Pincus}},
  \bibinfo{journal}{Phys. Rev. B} \textbf{\bibinfo{volume}{11}},
  \bibinfo{pages}{2042} (\bibinfo{year}{1975}).

\bibitem[{\citenamefont{Schulz}(1996)}]{Schulz1996}
\bibinfo{author}{\bibfnamefont{H.~J.} \bibnamefont{Schulz}},
  \bibinfo{journal}{Phys. Rev. Lett.} \textbf{\bibinfo{volume}{77}},
  \bibinfo{pages}{2790} (\bibinfo{year}{1996}).

\bibitem[{\citenamefont{Irkhin and Katanin}(2000)}]{IrkhinK2000}
\bibinfo{author}{\bibfnamefont{V.~Y.} \bibnamefont{Irkhin}} \bibnamefont{and}
  \bibinfo{author}{\bibfnamefont{A.~A.} \bibnamefont{Katanin}},
  \bibinfo{journal}{Phys. Rev. B} \textbf{\bibinfo{volume}{61}},
  \bibinfo{pages}{6757} (\bibinfo{year}{2000}).

\bibitem[{\citenamefont{Yasuda et~al.}(2005)\citenamefont{Yasuda, Todo,
  Hukushima, Alet, Keller, Troyer, and Takayama}}]{YasudaTHAKTT2005}
\bibinfo{author}{\bibfnamefont{C.}~\bibnamefont{Yasuda}},
  \bibinfo{author}{\bibfnamefont{S.}~\bibnamefont{Todo}},
  \bibinfo{author}{\bibfnamefont{K.}~\bibnamefont{Hukushima}},
  \bibinfo{author}{\bibfnamefont{F.}~\bibnamefont{Alet}},
  \bibinfo{author}{\bibfnamefont{M.}~\bibnamefont{Keller}},
  \bibinfo{author}{\bibfnamefont{M.}~\bibnamefont{Troyer}}, \bibnamefont{and}
  \bibinfo{author}{\bibfnamefont{H.}~\bibnamefont{Takayama}},
  \bibinfo{journal}{Phys. Rev. Lett.} \textbf{\bibinfo{volume}{94}},
  \bibinfo{pages}{217201} (\bibinfo{year}{2005}).

\bibitem[{\citenamefont{Todo}(2006)}]{Todo2006}
\bibinfo{author}{\bibfnamefont{S.}~\bibnamefont{Todo}}, \bibinfo{journal}{Phys.
  Rev. B} \textbf{\bibinfo{volume}{74}}, \bibinfo{pages}{104415}
  (\bibinfo{year}{2006}).

\bibitem[{\citenamefont{Hastings and Mudry}(2006)}]{HastingsM2006}
\bibinfo{author}{\bibfnamefont{M.~B.} \bibnamefont{Hastings}} \bibnamefont{and}
  \bibinfo{author}{\bibfnamefont{C.}~\bibnamefont{Mudry}},
  \bibinfo{journal}{Phys. Rev. Lett.} \textbf{\bibinfo{volume}{96}},
  \bibinfo{pages}{027215} (\bibinfo{year}{2006}).

\bibitem[{\citenamefont{Yao and Sandvik}(2007)}]{YaoS2007}
\bibinfo{author}{\bibfnamefont{D.~X.} \bibnamefont{Yao}} \bibnamefont{and}
  \bibinfo{author}{\bibfnamefont{A.~W.} \bibnamefont{Sandvik}},
  \bibinfo{journal}{Phys. Rev. B} \textbf{\bibinfo{volume}{75}},
  \bibinfo{pages}{052411} (\bibinfo{year}{2007}).

\bibitem[{\citenamefont{Sandvik}(1999)}]{Sandvik1999}
\bibinfo{author}{\bibfnamefont{A.~W.} \bibnamefont{Sandvik}},
  \bibinfo{journal}{Phys. Rev. Lett.} \textbf{\bibinfo{volume}{83}},
  \bibinfo{pages}{3069} (\bibinfo{year}{1999}).

\bibitem[{\citenamefont{Weiss}(1907)}]{Weiss1907}
\bibinfo{author}{\bibfnamefont{P.}~\bibnamefont{Weiss}}, \bibinfo{journal}{J.
  Phys. Th\'eor. Appl.} \textbf{\bibinfo{volume}{6}}, \bibinfo{pages}{661}
  (\bibinfo{year}{1907}).

\bibitem[{\citenamefont{Bethe}(1935)}]{Bethe1935}
\bibinfo{author}{\bibfnamefont{H.~A.} \bibnamefont{Bethe}},
  \bibinfo{journal}{Proc. Roy. Soc. (London)} \textbf{\bibinfo{volume}{A150}},
  \bibinfo{pages}{552} (\bibinfo{year}{1935}).

\bibitem[{\citenamefont{Peierls}(1936)}]{Peierls1936}
\bibinfo{author}{\bibfnamefont{R.}~\bibnamefont{Peierls}},
  \bibinfo{journal}{Proc. R. Soc. (London)} \textbf{\bibinfo{volume}{A154}},
  \bibinfo{pages}{207} (\bibinfo{year}{1936}).

\bibitem[{\citenamefont{Suzuki and Katori}(1986)}]{SuzukiK1986}
\bibinfo{author}{\bibfnamefont{M.}~\bibnamefont{Suzuki}} \bibnamefont{and}
  \bibinfo{author}{\bibfnamefont{M.}~\bibnamefont{Katori}},
  \bibinfo{journal}{J. Phys. Soc. Jpn.} \textbf{\bibinfo{volume}{55}},
  \bibinfo{pages}{1} (\bibinfo{year}{1986}).

\bibitem[{\citenamefont{Evertz}(2003)}]{Evertz2003}
\bibinfo{author}{\bibfnamefont{H.~G.} \bibnamefont{Evertz}},
  \bibinfo{journal}{Adv. in Physics} \textbf{\bibinfo{volume}{52}},
  \bibinfo{pages}{1} (\bibinfo{year}{2003}).

\bibitem[{\citenamefont{Todo and Kato}(2001)}]{TodoK2001}
\bibinfo{author}{\bibfnamefont{S.}~\bibnamefont{Todo}} \bibnamefont{and}
  \bibinfo{author}{\bibfnamefont{K.}~\bibnamefont{Kato}},
  \bibinfo{journal}{Phys. Rev. Lett.} \textbf{\bibinfo{volume}{87}},
  \bibinfo{pages}{047203} (\bibinfo{year}{2001}).

\bibitem[{\citenamefont{Tyablikov}(1959)}]{Tyablikov1959}
\bibinfo{author}{\bibfnamefont{S.~V.} \bibnamefont{Tyablikov}},
  \bibinfo{journal}{Ukrain. Mat. Zh.} \textbf{\bibinfo{volume}{11}},
  \bibinfo{pages}{287} (\bibinfo{year}{1959}).

\bibitem[{\citenamefont{Oguchi and Honma}(1963)}]{OguchiH1963}
\bibinfo{author}{\bibfnamefont{T.}~\bibnamefont{Oguchi}} \bibnamefont{and}
  \bibinfo{author}{\bibfnamefont{A.}~\bibnamefont{Honma}}, \bibinfo{journal}{J.
  Appl. Phys.} \textbf{\bibinfo{volume}{34}}, \bibinfo{pages}{1153}
  (\bibinfo{year}{1963}).

\bibitem[{\citenamefont{Stauffer and Aharony}(1994)}]{StaufferA1994}
\bibinfo{author}{\bibfnamefont{D.}~\bibnamefont{Stauffer}} \bibnamefont{and}
  \bibinfo{author}{\bibfnamefont{A.}~\bibnamefont{Aharony}},
  \emph{\bibinfo{title}{Introduction to Percolation Theory}}
  (\bibinfo{publisher}{Taylor \& Francis}, \bibinfo{address}{London},
  \bibinfo{year}{1994}), \bibinfo{edition}{2nd} ed.

\bibitem[{\citenamefont{Deng and Bl\"ote}(2005)}]{DengB2005}
\bibinfo{author}{\bibfnamefont{Y.}~\bibnamefont{Deng}} \bibnamefont{and}
  \bibinfo{author}{\bibfnamefont{H.~W.~J.} \bibnamefont{Bl\"ote}},
  \bibinfo{journal}{Phys. Rev. E} \textbf{\bibinfo{volume}{72}},
  \bibinfo{pages}{016126} (\bibinfo{year}{2005}).

\bibitem[{\citenamefont{Kato et~al.}(2000)\citenamefont{Kato, Todo, Harada,
  Kawashima, Miyashita, and Takayama}}]{KatoTHKMT2000}
\bibinfo{author}{\bibfnamefont{K.}~\bibnamefont{Kato}},
  \bibinfo{author}{\bibfnamefont{S.}~\bibnamefont{Todo}},
  \bibinfo{author}{\bibfnamefont{K.}~\bibnamefont{Harada}},
  \bibinfo{author}{\bibfnamefont{N.}~\bibnamefont{Kawashima}},
  \bibinfo{author}{\bibfnamefont{S.}~\bibnamefont{Miyashita}},
  \bibnamefont{and} \bibinfo{author}{\bibfnamefont{H.}~\bibnamefont{Takayama}},
  \bibinfo{journal}{Phys. Rev. Lett.} \textbf{\bibinfo{volume}{84}},
  \bibinfo{pages}{4204} (\bibinfo{year}{2000}).

\bibitem[{\citenamefont{Stinchcombe}(1979)}]{Stinchcombe1979}
\bibinfo{author}{\bibfnamefont{R.~B.} \bibnamefont{Stinchcombe}},
  \bibinfo{journal}{J. Phys. C: Solid State Phys.}
  \textbf{\bibinfo{volume}{12}}, \bibinfo{pages}{4533} (\bibinfo{year}{1979}).

\bibitem[{\citenamefont{Rushbrooke et~al.}(1972)\citenamefont{Rushbrooke, Muse,
  Stephenson, and Pirnie}}]{RushbrookeMSP1972}
\bibinfo{author}{\bibfnamefont{G.~S.} \bibnamefont{Rushbrooke}},
  \bibinfo{author}{\bibfnamefont{R.~A.} \bibnamefont{Muse}},
  \bibinfo{author}{\bibfnamefont{R.~L.} \bibnamefont{Stephenson}},
  \bibnamefont{and} \bibinfo{author}{\bibfnamefont{K.}~\bibnamefont{Pirnie}},
  \bibinfo{journal}{J. Phys. C: Solid State Phys.}
  \textbf{\bibinfo{volume}{10}}, \bibinfo{pages}{3371} (\bibinfo{year}{1972}).

\bibitem[{\citenamefont{Cheong et~al.}(1991)\citenamefont{Cheong, Cooper, Rupp,
  Batlogg, Thompson, and Fisk}}]{CheongCRB1991}
\bibinfo{author}{\bibfnamefont{S.-W.} \bibnamefont{Cheong}},
  \bibinfo{author}{\bibfnamefont{A.~S.} \bibnamefont{Cooper}},
  \bibinfo{author}{\bibfnamefont{L.~W.} \bibnamefont{Rupp}},
  \bibinfo{author}{\bibfnamefont{B.}~\bibnamefont{Batlogg}},
  \bibinfo{author}{\bibfnamefont{J.~D.} \bibnamefont{Thompson}},
  \bibnamefont{and} \bibinfo{author}{\bibfnamefont{Z.}~\bibnamefont{Fisk}},
  \bibinfo{journal}{Phys. Rev. B} \textbf{\bibinfo{volume}{44}},
  \bibinfo{pages}{9739} (\bibinfo{year}{1991}).

\bibitem[{LOO()}]{LOOPERweb}
\emph{\bibinfo{title}{{\tt http://wistaria.comp-phys.org/alps-looper/}}}.

\bibitem[{\citenamefont{Albuquerque et~al.}(2007)\citenamefont{Albuquerque,
  Alet, Corboz, Dayal, Feiguin, Fuchs, Gamper, Gull, G\"urtler, Honecker
  et~al.}}]{ALPS2007}
\bibinfo{author}{\bibfnamefont{A.}~\bibnamefont{Albuquerque}},
  \bibinfo{author}{\bibfnamefont{F.}~\bibnamefont{Alet}},
  \bibinfo{author}{\bibfnamefont{P.}~\bibnamefont{Corboz}},
  \bibinfo{author}{\bibfnamefont{P.}~\bibnamefont{Dayal}},
  \bibinfo{author}{\bibfnamefont{A.}~\bibnamefont{Feiguin}},
  \bibinfo{author}{\bibfnamefont{S.}~\bibnamefont{Fuchs}},
  \bibinfo{author}{\bibfnamefont{L.}~\bibnamefont{Gamper}},
  \bibinfo{author}{\bibfnamefont{E.}~\bibnamefont{Gull}},
  \bibinfo{author}{\bibfnamefont{S.}~\bibnamefont{G\"urtler}},
  \bibinfo{author}{\bibfnamefont{A.}~\bibnamefont{Honecker}},
  \bibnamefont{et~al.}, \bibinfo{journal}{J. Mag. Mag. Mat.}
  \textbf{\bibinfo{volume}{310}}, \bibinfo{pages}{1187} (\bibinfo{year}{2007}).

\end{thebibliography}

\end{document}